\documentclass[
    ,final            
  ]
  {aipproc}

\layoutstyle{6x9}

\begin{document}

\title{Doppler maps and surface differential rotation
of EI\,Eri from the MUSICOS 1998 observations}


\author{Zs.~K\H{o}v\'ari}{
  address={Konkoly Observatory, Budapest, Hungary}
}
\author{A.~Washuettl}{
  address={Astrophysikalisches Institut Potsdam, Germany}
}
\author{B.H.~Foing}{
  address={Research Support Division, ESA RSSD, ESTEC/SCI-SR, The Netherlands}
}
\author{K.~Vida}{
  address={E\"otv\"os University, Department of Astronomy, Budapest, Hungary},
altaddress={Konkoly Observatory, Budapest, Hungary}
}
\author{J.~Bartus}{
  address={Astrophysikalisches Institut Potsdam, Germany}
}
\author{K.~Ol\'ah}{
  address={Konkoly Observatory, Budapest, Hungary}
}
\author{the MUSICOS 98 team}{address=\empty
}

\begin{abstract}
We present time-series Doppler images of the rapidly-rotating active binary star
EI\,Eri from spectroscopic observations collected during the MUSICOS multi-site campaign in 1998, since the critical
rotation period of 1.947 days makes it impossible to obtain time-resolved images from a single site.
From the surface reconstructions a weak solar-type differential rotation, as well as a tiny poleward meridional
flow are measured.
\end{abstract}

\maketitle


\section{Introduction}

Due to the fast disperse of high resolution spectroscopic
observational facilities, surface differential rotation (hereafter DR)
measurements has become achievable for stars
of different types, giving a useful contribution for developing
solar and stellar dynamo theory. Indirect imaging techniques, such as Doppler
imaging can help in extending this input knowledge
by tracing the positions of different spots/magnetic features on consecutive surface maps.
The method of cross-correlating consecutive Doppler images to follow the time evolution of the spots
and to derive surface DR was demonstrated for the first
time for AB\,Dor \cite{1997MNRAS.291....1D}, and it has been used extensively thenceforth.
Unfortunately, rapid spot changes can fade out the correlation pattern of the DR.
On the other hand, averaging many cross-correlation maps can overwhelm the masking effect
of random-like short-term spot changes \cite{2004A&A...417.1047K}. 

In this paper we present time-series Doppler images for the rapidly rotating active
binary star EI\,Eridani (HD\,26337). From the consecutive surface maps we measure surface
DR. Since the rotation period of the star is nearly 2 days, data with
sufficient phase coverage for a Doppler image can only be collected over about 15 days from a single observing site. Thus, the reconstructed surface map is necessarily time-averaged,  
which makes DR measurements unfeasible. To avoid this problem, in our study we use
a set of high resolution spectra collected in the fifth MUSICOS (MUlti-SIte COntinuous Spectroscopy)
campaign in 1998, which allows to derive time-resolved surface maps.

\section{Observations}

\begin{figure}
 	\centering
 \includegraphics[width=0.5\textwidth]{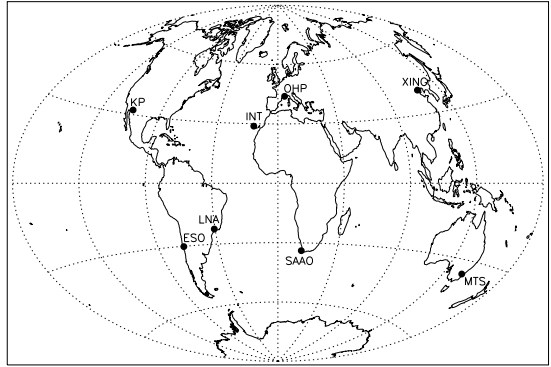}
 \caption{Observing sites involved in the MUSICOS 98 campaign}
 \label{music}
\end{figure}

The fifth MUSICOS campaign took place from November 22 to December 11, 1998.
It involved eight northern and southern
sites and ten telescopes of 2m class, mostly equipped with
cross-dispersed echelle spectrographs.
Fig.~\ref{music} shows the distribution of the participating
observatories.
From the total of 122 high-resolution spectra obtained in the campaign for
EI\,Eridani, 81 spectra were suitable for the purposes of Doppler imaging.
Spectroscopic observations are supported by simultaneous photometry
carried out with the Wolfgang-Amadeus twin Automatic Photoelectric Telescope (APT)
at Fairborn Observatory, Arizona \citep{kgs:boyd97}.
Phases of all line profiles are computed using the ephemeris
$HJD=2\,448\,054.7130 + 1.9472287 \times E$ \cite{wasi03}.

\section{Time-series Doppler imaging}

\begin{figure}
 	\centering
 \includegraphics[width=72mm]{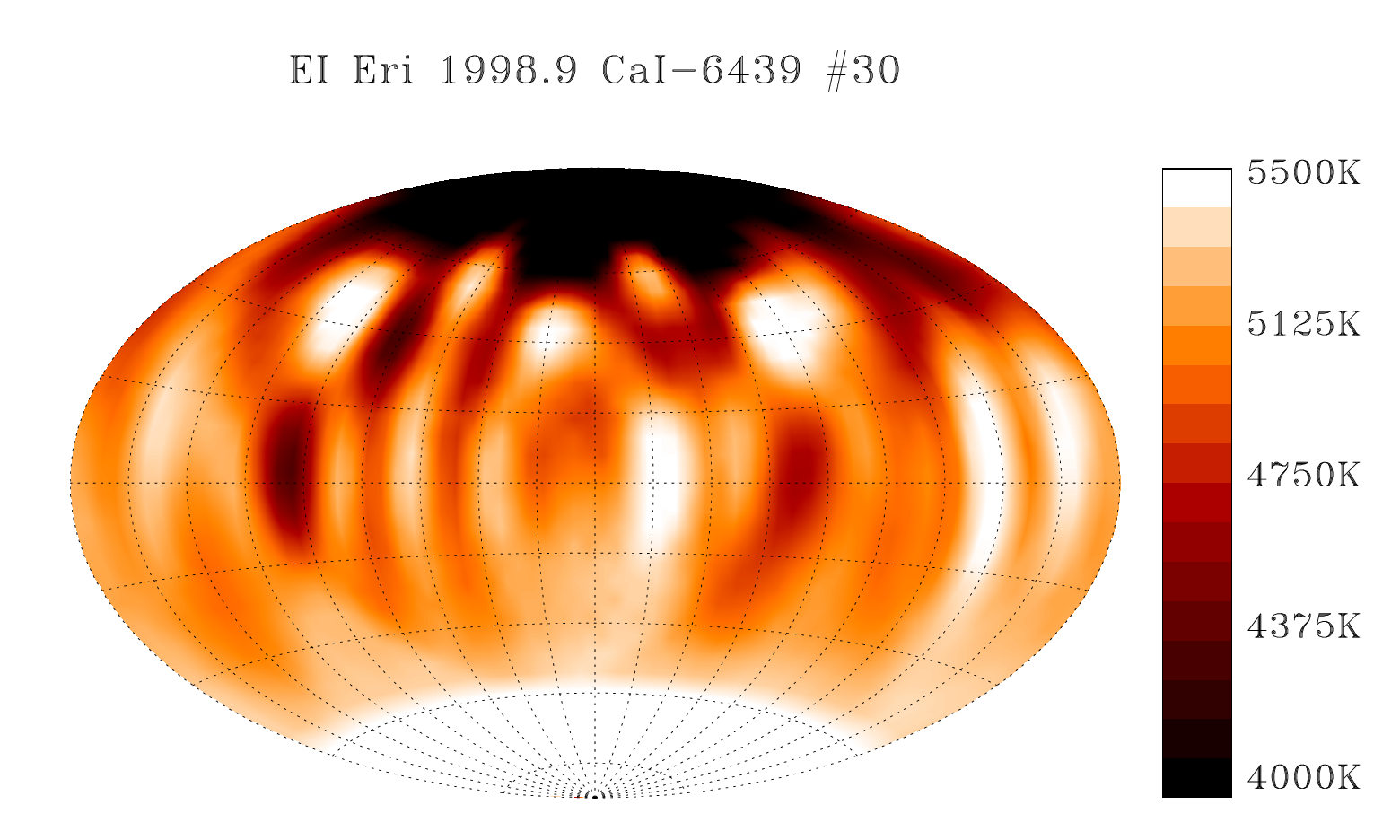}\includegraphics[width=72mm]{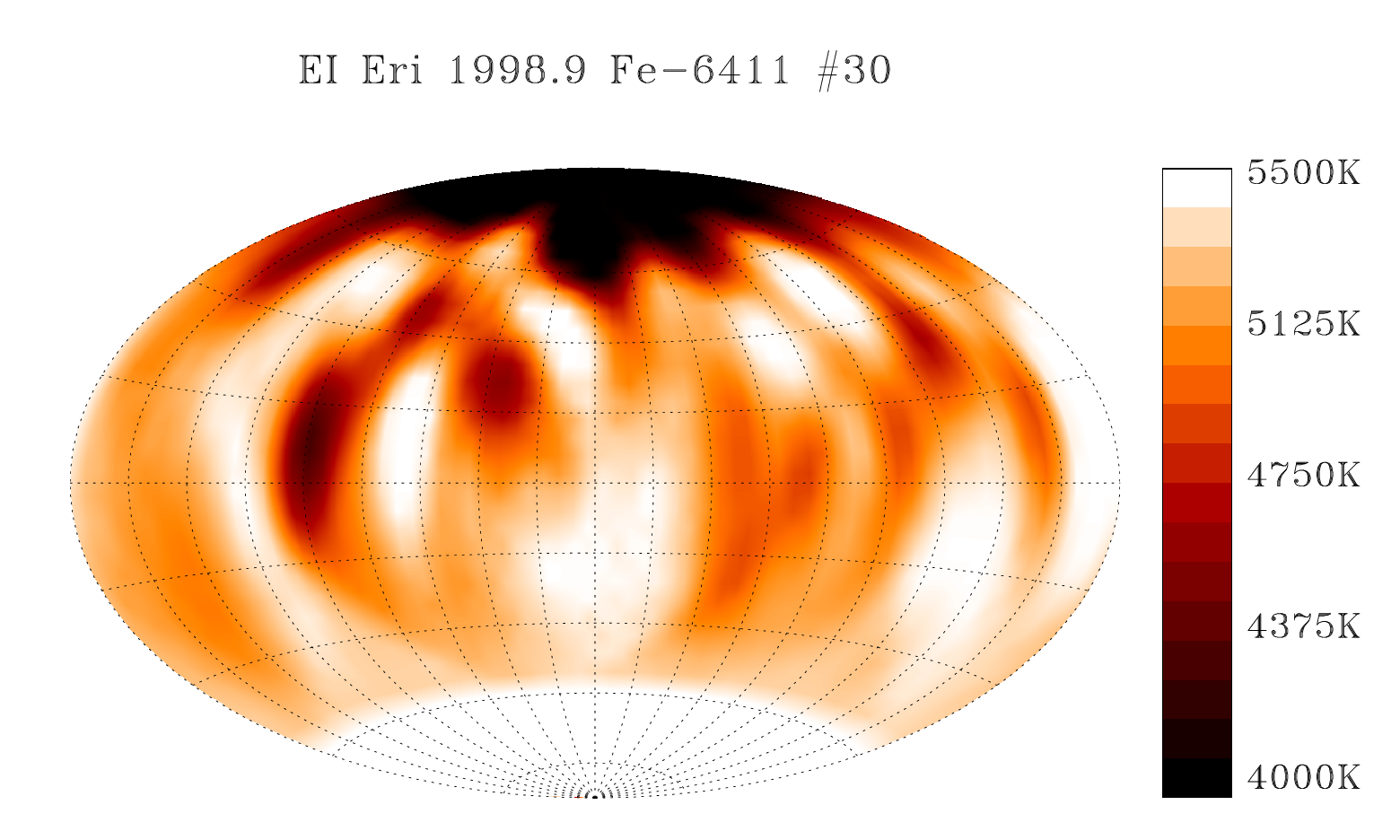}
 \caption{Simultaneous example image reconstructions from the time-series for the CaI-6439 (top)
and for the FeI-6411 (bottom) mapping lines}
 \label{dopplerimages}
\end{figure}

From the chronologically listed 81 spectra obtained during 20 days,
we form 52 subsets with 30-30 spectra. Each subset is formed from
the preceeding one by dropping the spectrum from the beginning of the list
and adding the subsequent one to the end. From these overlapping subsets we
compute 52 surface maps using our Doppler imaging code TempMap \cite{1989A&A...208..179R}.
As astrophysical input we use the data listed in Table~\ref{param} taken from
\cite{paper2}.
The resulted Doppler maps confirm the existence of a stable polar spot changing
slightly in size and shape, while low latitute spots are found to be short lived.
Independent inversions for the CaI-6439 and for the FeI-6411 mapping lines are
in good agreement, as seen in the corresponding example image pair in Fig.~\ref{dopplerimages}.

\begin{table}
\centering
 \begin{tabular}{ll}
  \hline
  \noalign{\smallskip}
  Parameter & Value     \\
  \noalign{\smallskip}
  \hline
  \noalign{\smallskip}
  $P_{\rm phot} = P_{\rm rot}$                  & 1.9472324\,d   \\
  T$_{\rm 0, phot} = $T$_{\rm 0, rot}$          & 2448054.7109\,d\\
  $\gamma$                                      & 21.64\,km/s   \\
  K1                                            & 26.83\,km/s   \\
  $e$ (eccentricity)                            & 0 (adopted)  \\
  $T_{\rm phot}$                                & 5500\,K       \\
  $T_{\rm max}$                                 & 6000\,K       \\
  $v\sin i$                                     & 51.0\,km/s    \\
  Inclination $i$                               & 56.0$^{\circ}$  \\
  $\log g$                                      & 3.5          \\
  Microturbulence $\xi$                        & 2.0\,km/s     \\
  Macroturbulence $\zeta_{\rm r} = \zeta_{\rm t}$ & 4.0\,km/s  \\
  $\log [Ca]$ abundance                                 & $-6.3$ (0.6\,dex below solar)\\
  $\log [Fe]$ abundance                                 & $-5.5$ (1.1\,dex below solar)\\
  \noalign{\smallskip}
  \hline
 \end{tabular}
\caption{Stellar parameters of EI\,Eri}\label{param}
\end{table}

\section{Differential rotation and meridional flow}

\begin{figure}
 \centering
 \includegraphics[width=72mm]{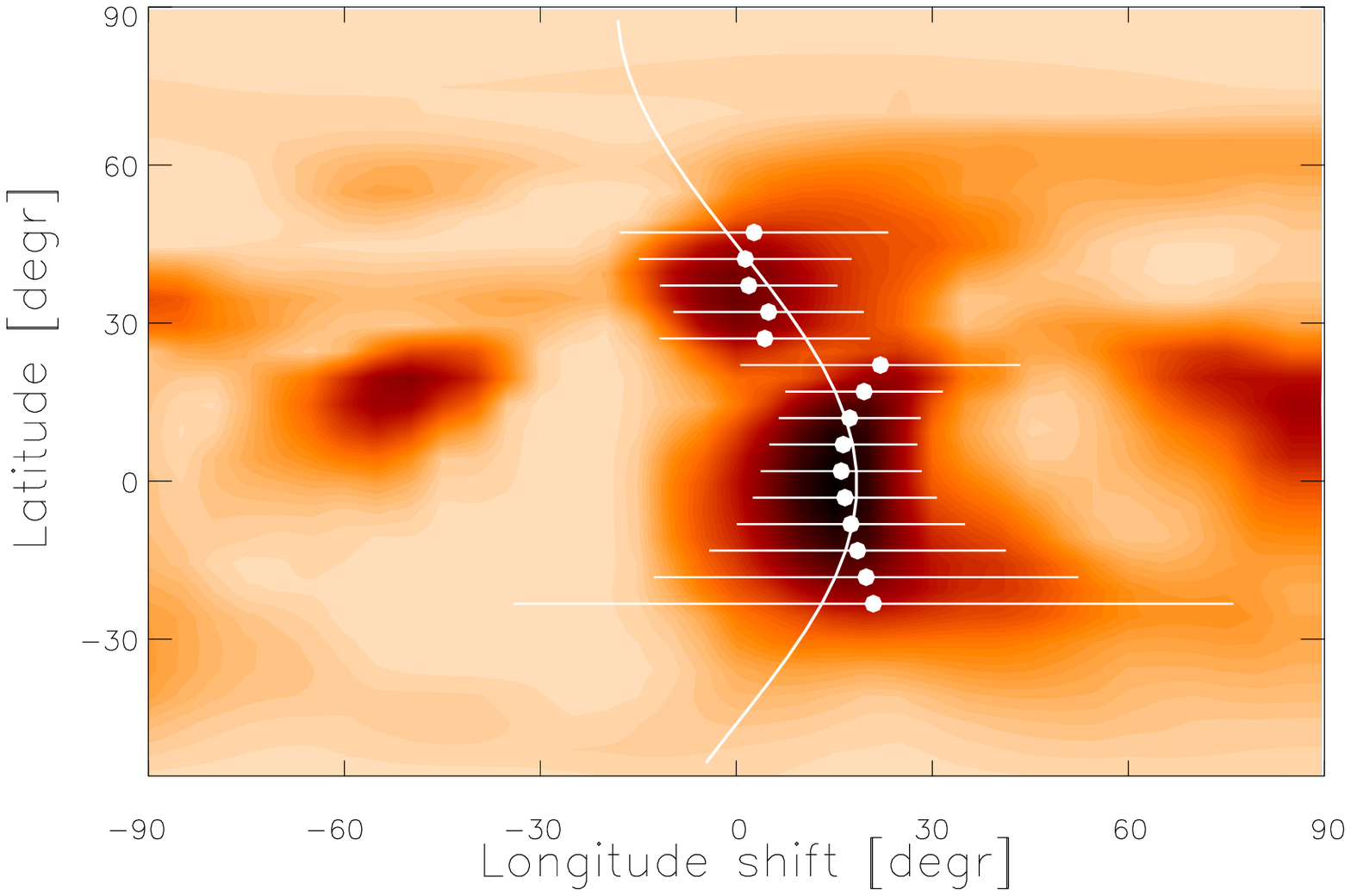}\includegraphics[width=72mm]{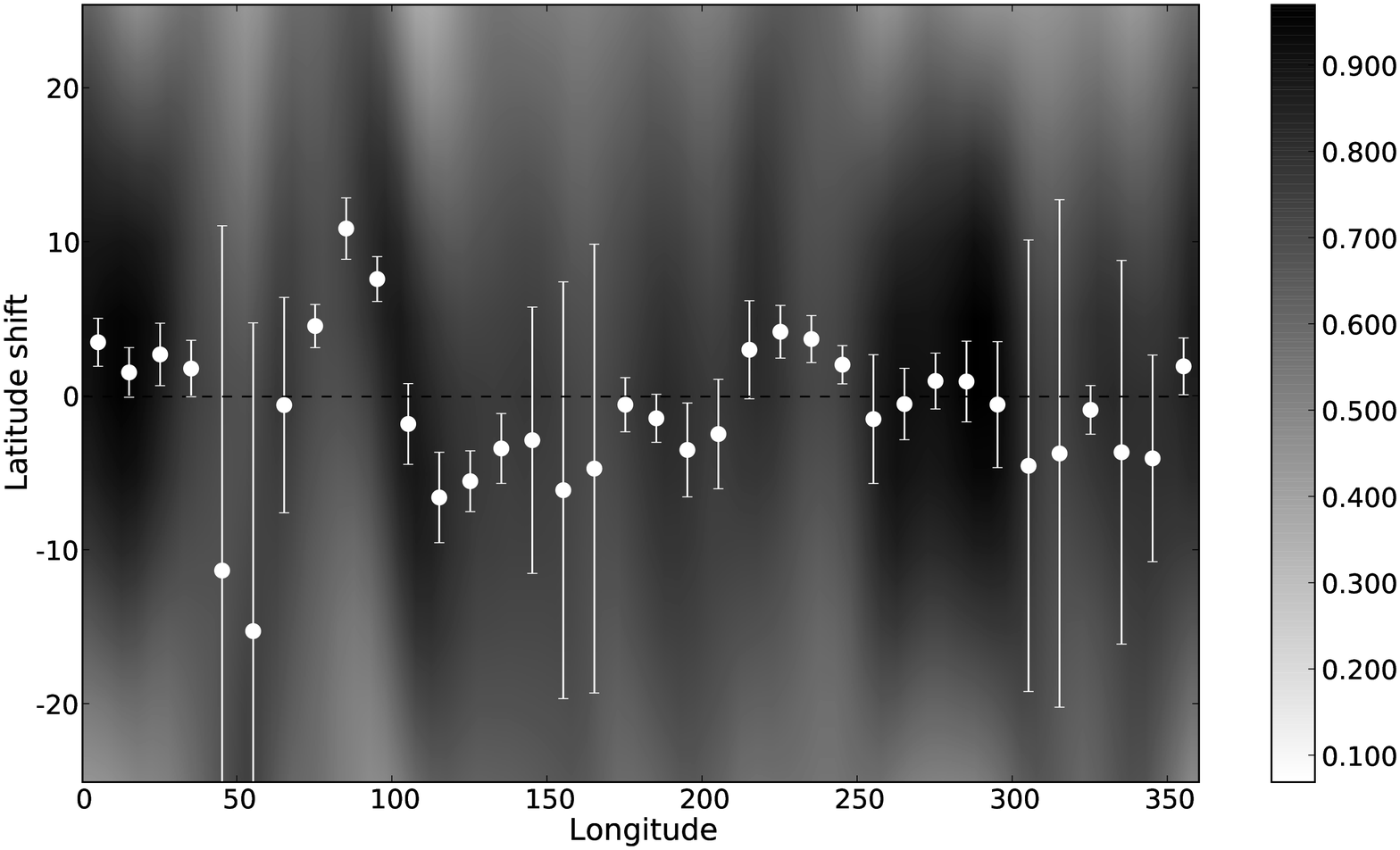}
 \caption{Resulting average ccf map with the best fit DR law (left) and the latitudinal ccf map (right)}
 \label{fig:ccf}
\end{figure}

For measuring surface DR we apply the technique called ACCORD
(Average Cross-CORrelation of time-series Doppler images)
introduced for the time-series Doppler maps of LQ\,Hya in \cite{2004A&A...417.1047K},
and applied more recently to $\sigma$\,Gem \cite{kovarietalsgem07} and to
$\zeta$\,And \cite{kovari:bartus:wasi07}.
The 52 time-series Doppler images are coupled in order to obtain 22 pairs of
 non-overlapping, consecutive Doppler maps. These pairs are cross-correlated
and the resulting correlation function (ccf) maps are averaged.
On the resulting average ccf map, we fit the correlation peak for each
latitudinal strip with a
Gaussian profile. The Gaussian peaks per latitude strips then represent
the DR pattern and can thus be fitted with a standard solar-like
quadratic DR law. The best fit DR function for the
combined Ca+Fe ccf map in the left panel of Fig.~\ref{fig:ccf} indicates a DR parameter
$\alpha=\Delta\Omega /\Omega$ of 0.037, about one-sixth of the value measured on the Sun.

An average meridional flow can be quantified by using ACCORD, since latitudinal motion of spots
can be analysed by cross-correlating the corresponding longitude
strips along the meridian circles. For this we use only the
hemisphere of the visible pole where Doppler
imaging is more reliable. The correlation maxima for each longitudinal strip are fitted
with a Gaussian and the resulting peaks represent the best
correlating latitudinal shifts. For a detailed description of the method see \cite{kovarietalsgem07}.
The latitudinal correlation pattern plotted in the right panel of Fig.~\ref{fig:ccf}
can be converted into an
average poleward velocity field of less than 100\,m/s.
For further details see our forthcoming paper \cite{forthcom}.

\begin{theacknowledgments}
ZsK is a grantee of the Bolyai J\'anos Scholarship of the Hungarian Academy of Sciences.
ZsK, AW, KV and KO are supported by the Hungarian Science Research Program (OTKA) grants
T-048961 and K 68626.

\end{theacknowledgments}

\bibliographystyle{aipprocl} 

\bibliography{mn-jour,kovaribib}

\end{document}